\documentclass[final,sort&compress
]{aipproc}
\layoutstyle{6x9}

\def\braket#1#2{\langle #1 |#2 \rangle}
\def\cm{{\cal M}}

\def\ep{\epsilon}

\begin{document}
\begin{flushright}
YITP-SB-02-78
\end{flushright}
\title{Two-loop QCD corrections for \\
$2 \to 2$ parton scattering processes\footnote{Based on talk given at the X
  Mexican School of Particles and Fields, Playa del Carmen, Mexico 2002}}
\author{Maria Elena Tejeda-Yeomans\footnote{
E-mail address: {\tt tejeda@insti.physics.sunysb.edu}. Partially supported 
by the National Science Foundation grant PHY-0098527.}}
{
address={C. N. Yang Institute for Theoretical Physics \\
State University of New York at Stony Brook, New York 11794-3840, USA}
}

\begin{abstract}
A summary is presented of the most recent matrix elements for massless $2 \to
2$ scattering processes calculated at two loops in QCD perturbation theory
together with a brief review on the calculational methods and techniques used.
\end{abstract}

\maketitle

\section{Introduction}
The description of high energy processes at hadron colliders together with
the study of the environment these events are emerging into, will rely
heavily on the predictive power of the theory of Quantum Chromo-Dynamics
(QCD). It is well known that QCD studies enable a 
better understanding of the color flow dynamics, which in turn facilitates 
improved predictions on observables such as $p_T$ distributions and
production rates. Also, QCD studies have a direct impact on theoretical
predictions for signals and their backgrounds, which are a key element in the
quest for {\it precision} physics at the highest energies. A plethora of data
on multi-particle final states will soon become available
from high-energy collider runs and the comparisons of jet observables with
theoretical predictions will be vital. In fact, missing higher order QCD
theoretical predictions for such observables may be large and
important. Therefore a systematic approach to perform these complex
calculations, which usually involve the analysis of multi-leg
and/or multi-loop amplitudes, becomes imperative. In recent years, new
techniques and integral manipulation methods have been
designed to achieve outstanding analytical results in this area. It is the
purpose of this talk to review briefly the latest results on two-loop QCD
corrections for massless $2 \to 2$ parton scattering processes and some of
the techniques used. This is not an exhaustive review, so more details can be
found in the references provided and the ones therein. 

There are many important reasons why one might consider a next-to-next to
leading order (NNLO) calculation\cite{Procs:2002} and to mention a few,
\begin{description}
\item[Reduced scale dependence]
The theoretical prediction for any observable $\Gamma$ should be independent
of the renormalisation scale $\mu$. But, if we perform a fixed 
order calculation a scale dependence is introduced. The change due to the 
variation of renormalisation scale is formally one order higher than the one
at which the theoretical prediction is given. Often, when giving a NLO
result, a variation of $\mu$ around some hard scale is used as a tool to give
the uncertainty in uncalculated higher orders but, this is an {\it estimate} 
of their size. The theoretical prediction can be improved with a complete
NNLO result. 
\item[Improved jet description]
If we consider higher order corrections we automatically improve the
matching between theoretically and experimentally defined jets. At leading
order we are modeling jets with single parton emission, but at higher orders
the phase space available is extended so that at NNLO up
to three partons can combine to form a jet. Corrections beyond LO
involve a better description of soft gluon radiation within
the jet and this may provide a more accurate picture of its shape and
structure. 
\item[Reduced power corrections]
When comparisons of NLO predictions with experimental data are
made, the need for power corrections (terms of ${\mathcal O}(1/Q^n)$)
arises. The structure of these corrections can be motivated theoretically but
they are always fitted to experiment. One might expect that higher order
corrections may play a role in reducing the size of these power
corrections.
\end{description}

\subsection{Partonic cross sections beyond NLO}
The description of hard scattering processes at future hadron colliders
requires the study of the factorized structure of a cross section for 
processes with quarks and gluons in the initial state. Up to power 
corrections, the inclusive factorized cross section can be written as
\begin{equation}
\sigma(P1,P2) \sim \sum_{ij} \int dx_1 dx_2 f_{i/1}(x_1,\mu_F^2)
f_{j/2}(x_1,\mu_F^2) \hat{\sigma}_{ij}(p_1,p_2,\alpha_s(\mu^2),s/\mu^2,s/\mu_F^2) 
\end{equation} 
where $p_i=x_i P_i$ is the momenta of the partons that initiate the hard
scattering and $f_{a/h}$ are the parton distribution functions
(pdf's). The non-perturbative effects are comprised in these distribution 
functions (and into fragmentation functions, in the exclusive case). 
$\hat{\sigma}$ is the hard scattering partonic cross section
which describes the interaction of partons $\{i,j\}$ that arose from hadrons
$\{1,2\}$ and can be calculated perturbatively, so that at NNLO for
2-particle production it can be written as
\begin{eqnarray}
\label{eq:hsmat}
\hspace{-0.5cm}\hat{\sigma}_{2~{\rm jet}} \sim&& 
   \int \Biggl[ |{ \braket{\cm^{(0)}}{\cm^{(0)}}}|^2\Biggr]_4 ~d\Phi_4 
+ \int
   \Biggl[ { \braket{\cm^{(0)}}{\cm^{(1)}}} + { \braket{\cm^{(1)}}{\cm^{(0)}}}
   \Biggr]_3 ~d\Phi_3 \nonumber \\ 
\lefteqn{\phantom{~}+\int \Biggl[{
   \braket{\cm^{(1)}}{\cm^{(1)}}} +{ \braket{\cm^{(0)}}{\cm^{(2)}}} + {
   \braket{\cm^{(2)}}{\cm^{(0)}}}\Biggr]_2 ~d\Phi_2}
\end{eqnarray}
where $[~]_n$ indicates the number of particles in the final state with
$d\Phi_n$ the corresponding phase space and $\cm^{(i)}$ the $i$-th order
scattering amplitude. After renormalisation, each of the integrals in 
Eq.~(\ref{eq:hsmat}) is ultra-violet (UV) finite but infra-red (IR) 
divergent which manifests itself as poles in $\ep$ (we adopt $D=4-2\ep$ as a
dimensional regulator)\footnote{The singularities are guaranteed to cancel
  for sufficiently inclusive physical quantities\protect{\cite{KLN}}.}. The
integration over phase space and the
cancellation of poles requires the study of the kinematical regions where the
additional radiated particles become unresolved\footnote{See for example,
analytical\protect{\cite{BabisMelnikov}} and numerical\protect{\cite{Gundrun:2002}} applications.}. Finally, pdf's and their
evolution are needed at an accuracy that matches that of the matrix element
calculation and for NNLO some great developments have already taken
place\cite{PDF:2002}.

\section{Two-loop integrals}
The number of diagrams involved in the complete two-loop and one-loop matrix
  elements needed for the NNLO contribution to inclusive jet cross sections
  and photo-production at hadron colliders, are shown in
  Table~\ref{tab:numofdiags} \cite{BGOTY}. 
\begin{table}
\begin{tabular}{c|c|c|c||c|c|c|c }
\hline
\multicolumn{8}{c}{Number of Diagrams} \\
 \hline 
Process  &   Tree & One loop & Two loops & Process  &   Tree & One loop & Two loops \\
\hline
$gg\to gg$ & 4 & $\sim 80$ & $\sim 1700 $ & $q_1 \bar q_1 \to q_2 \bar q_2$ & 1 & $\sim 10$ & $\sim 200 $ \\
$q \bar q \to gg$ & 3 & $\sim 30$ & $\sim 600 $  & $q \bar q \to g\gamma$ & 2 & $\sim 10$ & $\sim 300 $ \\
$q \bar q \to q \bar q$ & 2 & $\sim 20$ & $\sim 200 $ & $q \bar q \to \gamma\gamma$ & 2 & $\sim 10$ & $\sim 140 $ \\
\hline
\multicolumn{8}{c}{$+$ crossed processes} \\
\hline
\end{tabular}
\caption{Number of Feynman diagrams for different 
$2 \to 2$ scattering processes}
\label{tab:numofdiags}
\end{table}
The types of integrals arising in these matrix-element calculations include
integrals with scalar numerators
\begin{equation}
\label{eq:scalarint}
\int \frac{d^D k}{i \pi^{\frac{D}{2}}} \int \frac{d^D \ell}{i
  \pi^{\frac{D}{2}}} \frac{{\mathcal F}(k_i\cdot k_j, k_i\cdot p_j,p_i \cdot
  p_j)}{A^{\nu_1}_1 \cdots A^{\nu_n}_n},
\end{equation}
where $k_i = k, \ell$ are the loop momenta, $p_i$ are the external momenta
and ${\mathcal F}$ is a scalar function. Here, the massless 
propagators are denoted by $A_i\sim (k_i \pm \wp + i\varepsilon)^2$ 
where $\wp$ can be 
any of the loop or external momenta. There are other terms that would typically
look like the one shown in Eq.~(\ref{eq:scalarint}), except now the 
function ${\mathcal F} \to {\mathcal
  F}^{\mu \nu \cdots}(k,\ell)$ is a tensor that can depend on the loop-momenta of 
the system and/or other tensors (such as the metric tensor $g^{\mu \nu}$).
Furthermore, the denominator of an integral provides the momentum flow in the
graph and gives a complete description of its skeleton or topology which can 
be {\it planar} or {\it non-planar}. Within these two
categories we will identify sub-topologies or {\it pinchings}, depending on
whether or not a particular subset of propagators is absent.     
\begin{figure}
  \scalebox{.19}{\includegraphics{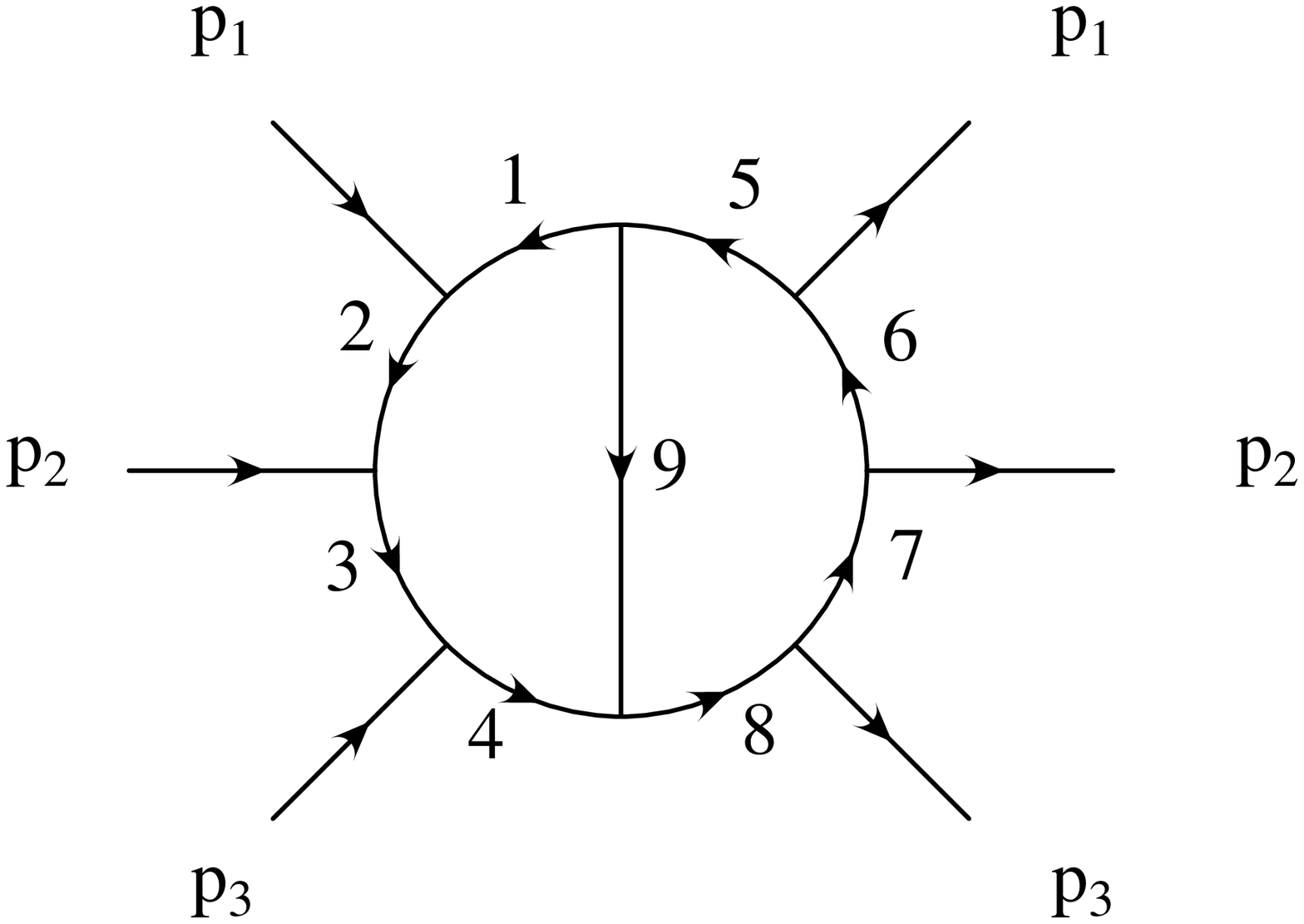}}
  \scalebox{.9}{\includegraphics*[4.5cm,10.5cm][14.5cm,14cm]{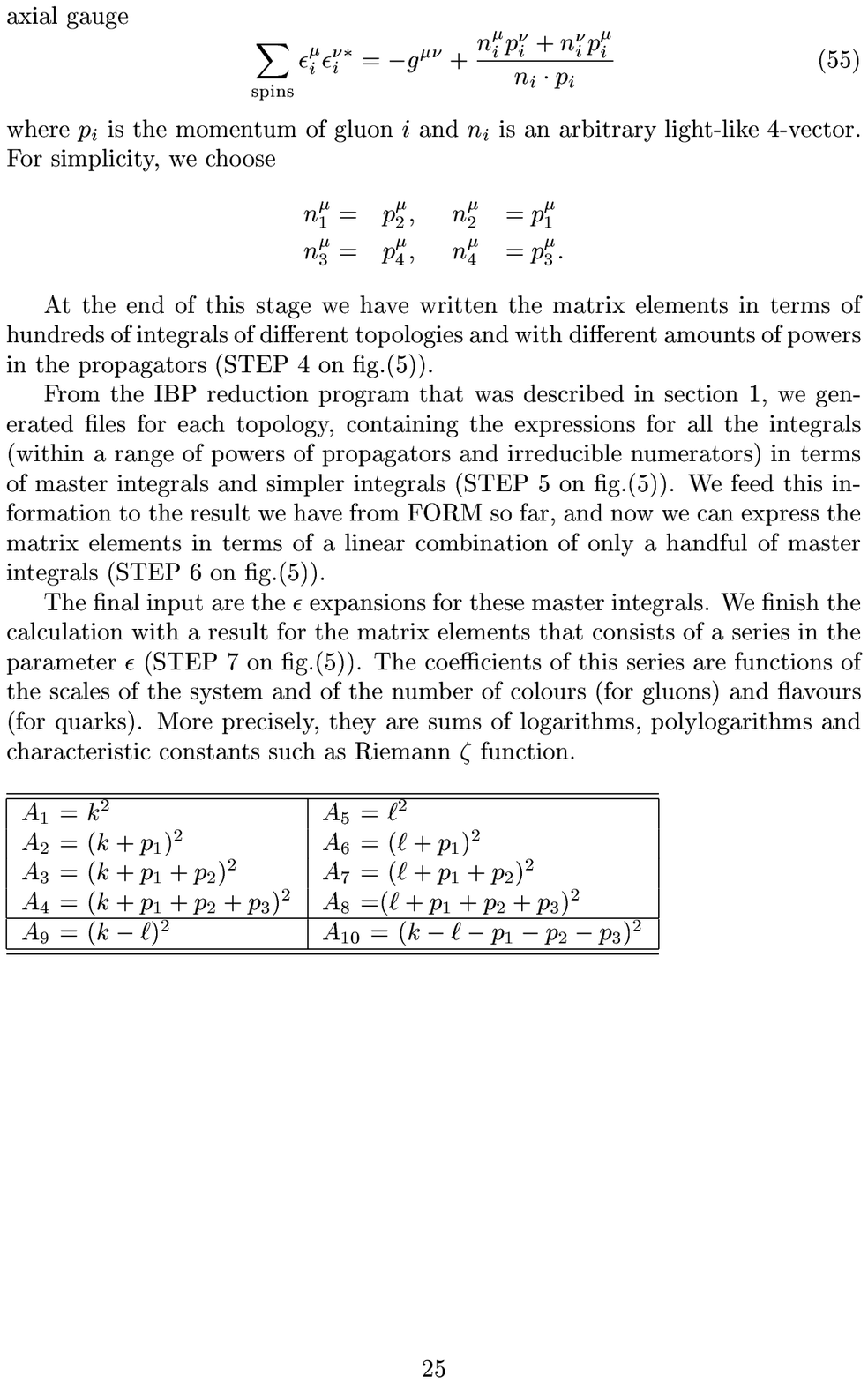}}
  \scalebox{.19}{\includegraphics{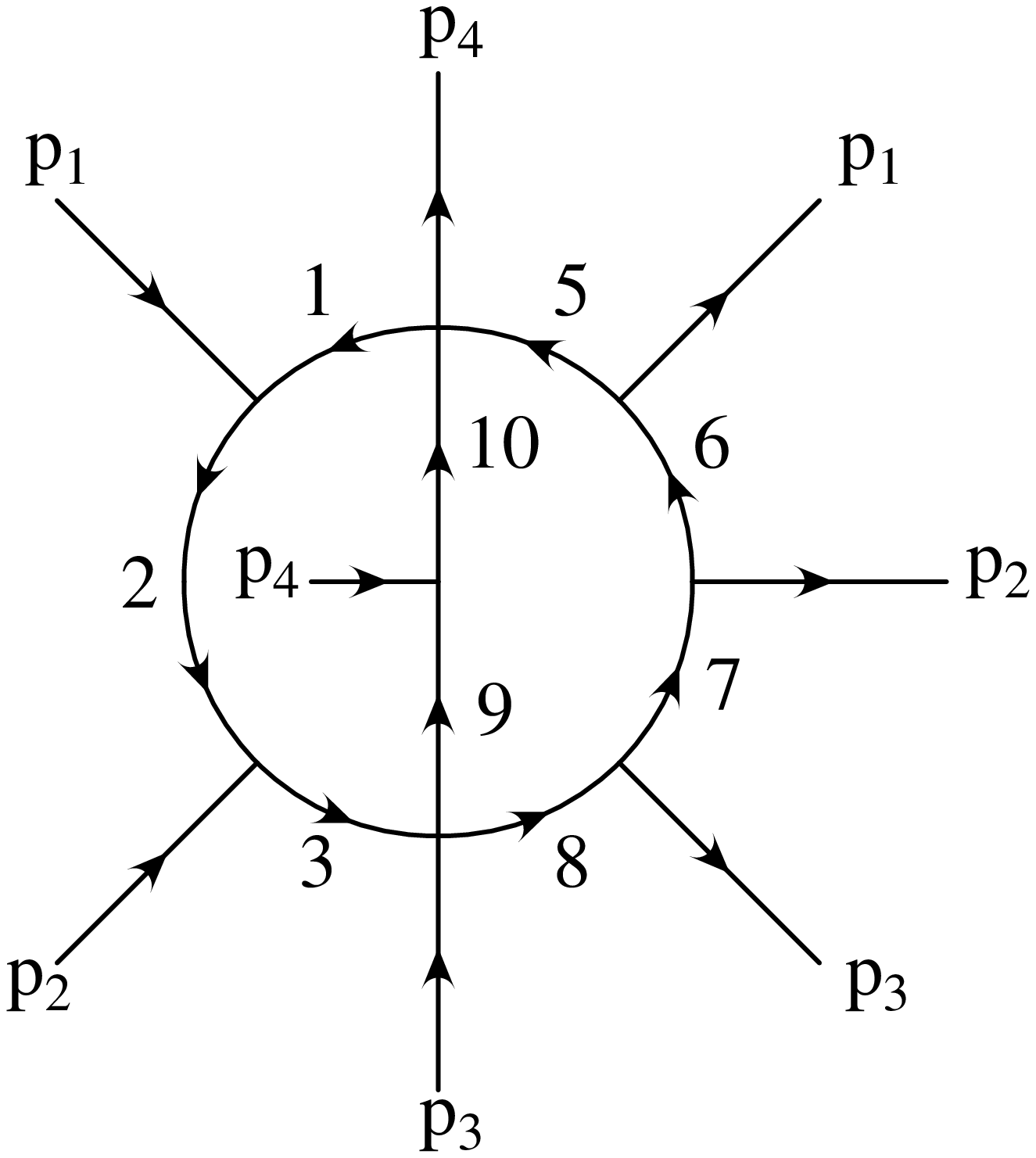}}
  \caption{Auxiliary diagrams and table with description of propagators}
  \label{fig:auxdiag}
\end{figure}
As technical aids, we use a couple of auxiliary general diagrams shown
in Fig.~\ref{fig:auxdiag}, where each of the propagators is labeled by an 
integer and carries a specific momentum that fulfills conservation 
of momenta throughout. This auxiliary representation allows us
to manipulate hundreds of integrals with the minimum amount of information
and without compromising the accuracy of the integral description. So to
refer to a general $D$-dimensional integral with 10 propagators raised to 
arbitrary powers, we use
$I^D[\nu_1,\nu_2,\nu_3,\nu_4,\nu_5,\nu_6,\nu_7,\nu_8,\nu_9,\nu_{10}]$. 
Any planar or non-planar topology is represented by eliminating, increasing or decreasing 
the values of $\nu_i$. 
\begin{figure}
  \scalebox{0.8}{\includegraphics*[4cm,13cm][21cm,18cm]{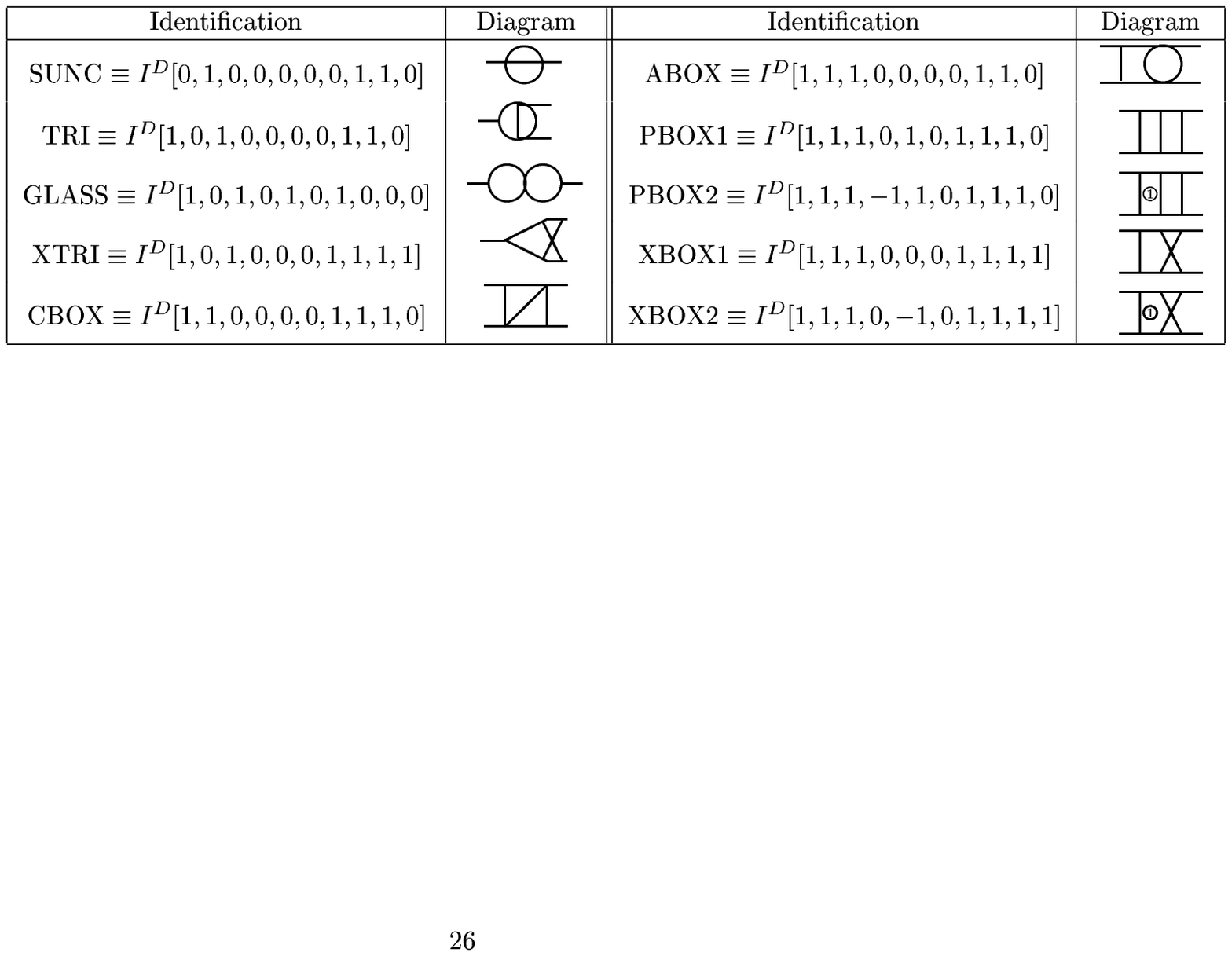}}
  \caption{Master integrals needed for the two-loop massless $2\to 2$ parton hard
    scattering matrix evaluation}
  \label{fig:master}
\end{figure}
It turns out that the two-loop QCD matrix-element calculation for $2 \to 2$
scattering processes with massless external states, requires the {\it explicit}
analytic evaluation of the 10 {\it master} integrals (MI) shown in 
Fig.~\ref{fig:master}. Any other integral that contains more powers in the 
propagators or higher dimensions can be written as a linear combination of 
the MI and below we will explore briefly how this can be done. First though, 
we will review the methods that are used to evaluate these MI explicitly.

\subsection{Explicit loop integration}
\label{sec:explicitint}
The strategically important analytic expressions for all MI have been
provided in Refs.\cite{SmirnovTausk,AlgorIBPLI,GehrmannRemiddi} as series in 
$\ep = (4-D)/2$. Usually, Feynman or Schwinger parameterizations (see
for example \cite{BOOKREF}) are used,
whereby the propagators of the loop integrand are expressed in terms of
integrations for real parameters over a particular range. In both of these 
approaches, the loop momenta integration can be done easily and the remaining 
integrations over the parameters are doable for reasonably sized topologies.
Sometimes, using these parametric representations does not leave an 
integral that can be solved easily, even more so when the number of loops, 
external legs and kinematical scales increase. Different techniques arose due 
to this situation, among others
\begin{itemize}
\item The Mellin-Barnes method is based on the representation for a
  sum to some power, as a contour integral over a complex variable and the
integration is then performed on straight contour lines parallel to the
imaginary axis. After closing a contour, the result is the sum of all
enclosed residues that may be expressed as a hypergeometric series. This has
been used to obtain vital results for the two-loop
boxes\cite{SmirnovTausk,Smirnov2}.
\item The Negative Dimensions technique consists of rewriting the integral 
over the parameters by introducing new ones through a multinomial expansion. 
Many conditions have to be satisfied among the parameters, which leads to 
the restriction: $D$ must be a negative integer. Some results for MI have
been obtained using this method\cite{AlgorIBPLI}. 
\item Numerical strategies are used, where any loop
  integral is stripped analytically of its IR singularities so that the
  finite integrals can be performed numerically. These methods
  can also be used to tackle the phase space integrations in the calculation 
of observables\citep{Gundrun:2002}.
\item The analytic evaluation of MI can also be carried out without
  explicit integration over loop momenta by deriving differential equations
  for MI in internal propagator masses or in external momenta. The equations 
can be solved with appropriate boundary conditions. This approach has been
widely used to evaluate MI, as was done recently for $e^+ e^- \to 3
~jets$\citep{GehrmannRemiddi}.
\end{itemize}

\subsection{Loop integral reduction}
This approach produces an environment in which complex topologies with high powers
on the propagators can be reduced down to integrals that can be solved with
the methods described in the previous section. This reduction can be achieved 
using systems of equations that stem from Integration by Parts identities
(IBP)\cite{IBPiden} and exploiting the Lorentz
invariance\cite{GehrmannRemiddi} of Feynman integrals. Let us review briefly
how these identities are generated and help reduce the complexity of an
integral. Consider a general two-loop integral
\begin{eqnarray}
I^D[\nu_1,...,\nu_n] &=& 
\int 
\frac{d^Dk}{i\pi^{\frac{D}{2}}}
\int
\frac{d^D\ell}{i\pi^{\frac{D}{2}}}
\frac{1}{A_1^{\nu_1} \cdots A_{n}^{\nu_{n}}},
\label{eq:loopint}
\end{eqnarray}
The idea behind IBP is to generate relations between loop integrals 
through a total derivative with vanishing surface terms, expressed 
as the following identity
\begin{eqnarray}
\label{eq:IBPid}
\int \frac { d^{D}k }{ i \pi^{\frac{D}{2}} } 
\int \frac { d^{D}\ell }{ i \pi^{\frac{D}{2}} }
\frac {\partial}{\partial k^{\mu}_{i}}
\left[
\frac { {\cal V}^{\mu} }{ A^{\nu_{1}}_{1}\cdots A^{\nu_{n}}_{n} } \right] 
& \equiv & 0, 
\end{eqnarray}
where $k_i=k,\ell$ are the loop-momenta and ${\cal V}$ can be any internal or 
external momenta involved in the loop integration. Executing the derivative
on all possible choices  for ${\cal V}^{\mu}$ 
will generate a set of relations\footnote{For a graph with $m$ loops and  $n$
  independent external momenta, we can generate $m(m+n)$
  identities.} between integrals with dot
products in the numerator. These dot products can be rewritten in terms of
linear combinations of propagators, by means of relations such as
\begin{eqnarray}
\label{eq:dotprod}
2(k + g)\cdot(k + h) & = & (k + g)^{2} + (k + h)^{2} - (g - h)^{2}.
\end{eqnarray}
With this simple step we can rewrite all the contents of the numerator
in terms of propagators that may or may not be part of the denominator. We
then say that the numerator is {\it reducible} if we can cancel it through
and {\it irreducible} otherwise. In most cases, we can exchange the problem of 
calculating the original integral, for the problem of calculating 
a set of simpler integrals. We can imagine studying all the IBP identities
for the topologies involved in a particular matrix-element calculation
and applying the reduction procedure iteratively. Then we can assemble an
algorithm that takes any integral and expresses it in terms of a few
MI. This is precisely what we will review next.

\section{Matrix element evaluation and results}
\begin{figure}
  \scalebox{.9}{\includegraphics*[4.2cm,8cm][18.2cm,20.8cm]{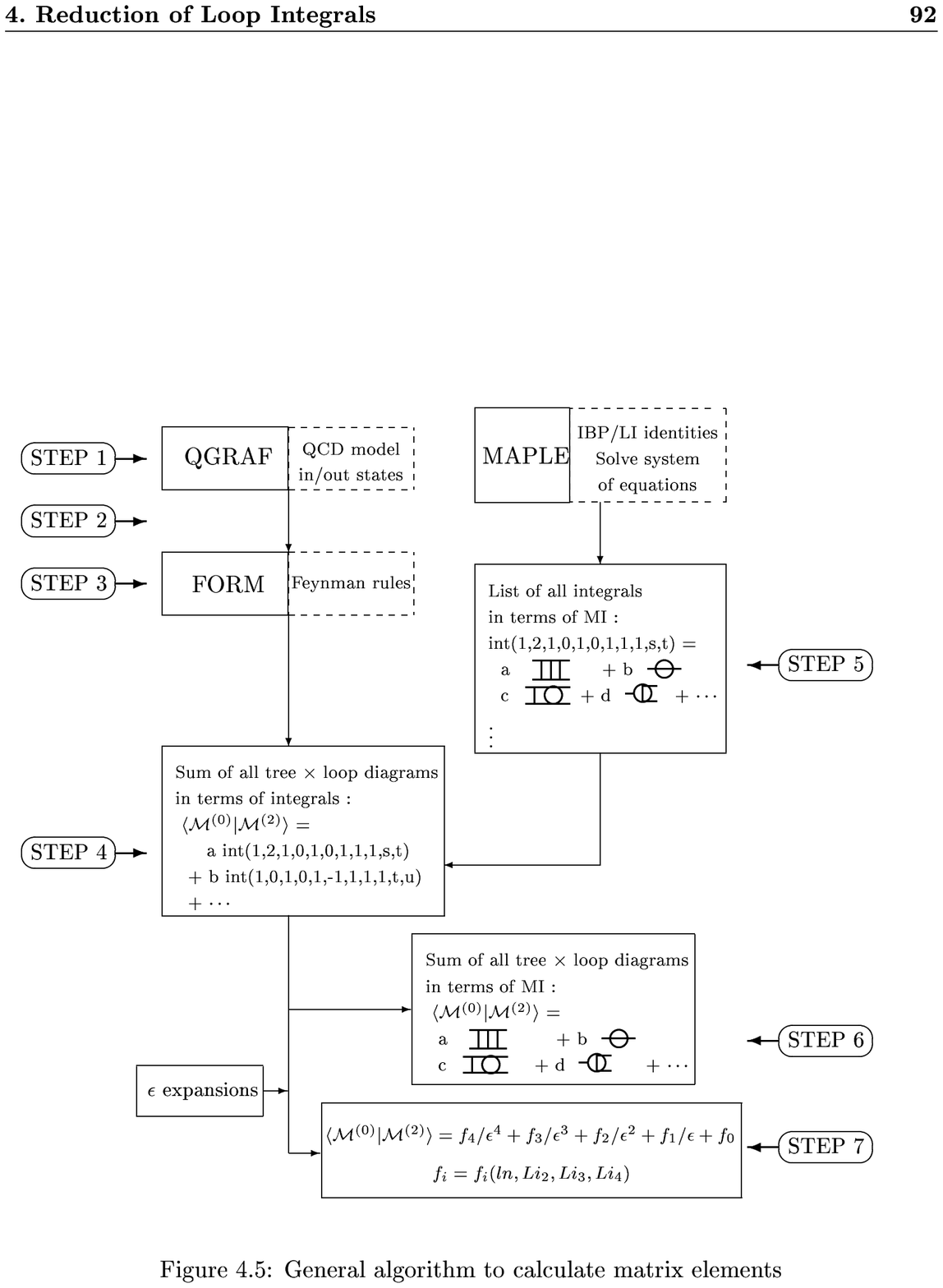}}
  \caption{Basic steps to follow in the evaluation of matrix elements}
  \label{fig:matelechart}
\end{figure}

The ideas exposed above and extensions to them have been used in the past few
years to tackle two-loop calculations. The way these calculations are put
together varies, for example the chart in Fig.~\ref{fig:matelechart} shows 
some basic steps we followed to evaluate the two-loop QCD matrix-elements 
for $2 \to 2$ scattering.
Steps 1 through 3 are mainly related to the automatic generation of diagrams
using QGRAF\cite{QGRAF} and the tensor algebra manipulation using
FORM\citep{FORM}. By step 4, we have the matrix element written as a
linear combination of many different scalar and tensor two-loop
integrals. These integrals can be rewritten in terms of MI (step 6), using a
reduction system that was generated using MAPLE (step 5). The final
step is the input of $\ep$-expansions for each MI, so that we get
\begin{equation}
\label{eq:polesme}
\braket{\cm^{(0)}}{\cm^{(2)}} \sim \sum_{n=0}^{4} \frac{f_n\left( 
\alpha_s,\mu,\{ s,t,u \} \right)}{\ep^n},
\end{equation}
where $\{ s,t,u \}$ are the usual Mandelstam variables. The functions
$f_n$ depend on the color factors $\{N_F,C_F, C_A, T_R\}$ and logarithms of 
ratios of the kinematic variables. The finite piece $f_0$ can contain up to 6
polylogarithms (${\rm Li}_n$ with
$n=\{2,3,4\}$) of ratios in the kinematical variables, and the Riemann zeta 
function $\zeta(n)$ (with $n=\{2,3\})$.

Several checks have served to verify the results obtained with this approach
such as comparisons with QED processes\citep{Bern1} and also the study of 
the analytic structure of QCD amplitudes in the limit of forward and backward 
scattering together with the confirmation of the two-loop gluon Regge 
trajectory\citep{DelDuca}. The fact that the singular analytical structure 
in Eq.~(\ref{eq:polesme}) can be obtained independently, is also a powerful 
check on the calculation\cite{BGOTY}. Catani\citep{Catani} proposed the structure of 
the $1/\ep^n ~\forall ~n=\{2,3,4\}$ poles together with the color uncorrelated
structure of the $1/\ep$ pole. For a while, the origins of this proposal did
not exist. However, in the work with Sterman\citep{Sterman}, we 
describe how the factorization properties of loop amplitudes lead to the
exponentiation of double and single poles at each order in perturbation
theory. The poles can then be assembled in terms of universal functions 
associated with the external partons. This formalism provides a way to
generate the complete pole structure for multi-loop amplitudes at two loops
and beyond.

\section{Conclusion and outlook}
As reviewed, the past few years have seen a breakthrough in multi-loop 
integration technology and outstanding progress in the calculation of 
two-loop matrix elements in QCD. Much work remains to be done to have NNLO
Monte Carlo numerical estimates but, looking back at the results we now have,
it seems they will be available soon for the first basic scattering
processes. This will enable an improved description of high-energy QCD phenomena.

\begin{theacknowledgments}
I would like to thank the organizers of the X Mexican School of Particles and
Fields for creating a stimulating and productive environment. I am happy to
thank Babis Anastasiou, Nigel Glover, Carlo Oleari, George Sterman and Bas
Tausk for very stimulating collaborations. This work was partially supported
by the National Science Foundation grant PHY-0098527.
\end{theacknowledgments}

\end{document}